\documentclass[conference]{IEEEtran}
\IEEEoverridecommandlockouts
\usepackage{cite}
\usepackage{amsmath,amssymb,amsfonts}
\usepackage{algorithmic}
\usepackage{graphicx}
\usepackage{textcomp}
\usepackage{xcolor}
\def\BibTeX{{\rm B\kern-.05em{\sc i\kern-.025em b}\kern-.08em
    T\kern-.1667em\lower.7ex\hbox{E}\kern-.125emX}}

\usepackage{authblk}

\usepackage{caption}
\usepackage{booktabs}
\usepackage{graphicx}
\usepackage{colortbl}
\usepackage{amsmath}
\usepackage{mathtools}
\usepackage{diagbox}
\usepackage{pgfplots, pgfplotstable}
\usepackage{multirow}
\pgfplotsset{compat=1.11,
    /pgfplots/ybar legend/.style={
    /pgfplots/legend image code/.code={%
       \draw[##1,/tikz/.cd,yshift=-0.25em]
        (0cm,0cm) rectangle (3pt,0.8em);},
   },
}
\usepgfplotslibrary{colormaps}
    \def\addlegendimage{\csname pgfplots@addlegendimage\endcsname}

\pgfplotsset{ 
cycle list={%
{draw=black,mark=star,solid},
{draw=black, mark=square,solid}}}

\usepackage{url}

\usepackage{fancyhdr}
\fancypagestyle{firstpage}
{
    \fancyhead[L]{\footnotesize \textcopyright 2026 IEEE.  Personal use of this material is permitted. Permission from IEEE must be obtained for all other uses, in any current or future media, including reprinting/republishing this material for advertising or promotional purposes, creating new collective works, for resale or redistribution to servers or lists, or reuse of any copyrighted component of this work in other works. The paper is accepted at IEEE IOLTS'26.}
    \fancyhead[R]{}
}

\begin{document}


\title{\LARGE \textbf{Effective and Memory-Efficient Alternatives to ECC for Reliable Large-Scale DNNs}}

\author[1]{Mohammad Hasan Ahmadilivani}
\author[1]{Marten Roots}
\author[2]{Marco Restifo}
\author[1]{\\Sven-Markus Loorits}
\author[2]{Luca Di Mauro}
\author[1]{Jaan Raik}
\affil[1]{Tallinn University of Technology, Tallinn, Estonia}
\affil[2]{ARM Ltd., Cambridge, United Kingdom}
\affil[1]{\{mohammad.ahmadilivani, jaan.raik\}@taltech.ee}
\affil[2]{\{marco.restifo, luca.dimauro\}@arm.com}












\maketitle
\thispagestyle{firstpage}
\small

\begin{abstract}

Modern Deep Learning (DL) workloads are increasingly deployed in safety-critical domains, such as automotive systems and hyperscale data centers, where transient hardware faults pose a serious threat to system reliability. These workloads are highly memory-intensive, and their correct functionality strongly depends on model parameters stored in memory, which are typically protected using Error Correction Codes (ECCs). In this work, we study ECC's impact on such models and propose two lightweight alternatives to ECCs that achieve superior reliability. The first approach, \textit{MSET}, selectively hardens the most vulnerable bits in CNN and ViT parameters, while the second approach, \textit{CEP}, provides fine-grained protection for all parameter bits. Experimental results demonstrate that both methods significantly enhance the reliability of large CNNs and ViTs, mostly outperforming conventional Single Error Detection Double Error Correction (SECDED) ECC schemes, with no memory overhead and, in fact, with considerably lower area and delay characteristics when compared to SECDEC. Experimental results indicate that ViTs can be effectively protected by merely protecting their highest exponent bits in FP16 and FP32 representations. Furthermore, applying the CEP technique can guarantee the resilience of DNNs by up to one order of magnitude higher BERs, with a $3.5\times$ lower area overhead and $7\times$ faster decoder compared to SECDED ECC.

\end{abstract}

\section{Introduction} \label{sec:intro}

Recent advances in Artificial Intelligence (AI) and Machine Learning (ML) have enabled achieving remarkable accuracy across a broad spectrum of applications, driving their rapid adoption in both edge and cloud environments \cite{liu2024emerging}. Contemporary ML models, such as large-scale Convolutional Neural Networks (CNNs) and Vision Transformers (ViTs), attain such accuracy at the cost of significant computational complexity and memory footprint. These growing resource demands place considerable pressure on heterogeneous computing platforms and intensify their vulnerability to hardware-induced faults. In this context, even infrequent errors occurring during computation or data movement can be amplified throughout the inference process, ultimately leading to incorrect or unreliable outputs \cite{ahmadilivani2024systematic,rech2024artificial}.

In safety-critical EdgeAI applications, such as autonomous vehicles, hardware faults can result in severe or even catastrophic consequences \cite{Accident}. Similarly, in High-Performance Computing (HPC) environments, a large number of faults manifest as Silent Data Corruptions (SDCs) on a daily basis, adversely affecting the reliable deployment of ML models on large-scale server infrastructures \cite{dixit2021silent, Nishant2025sdc}. These failure modes pose significant challenges to the correctness and robustness of ML-driven workloads.
The scale of modern DNNs further necessitates the need for fault-tolerance mechanisms in memories. State-of-the-art DNNs commonly comprise hundreds of millions to billions of parameters \cite{zhai2022scaling}, all of which are stored in DRAM and SRAM memories that are inherently susceptible to soft errors \cite{ibe2010impact}. Such low-level memory faults can silently corrupt model parameters, propagating to the output and ultimately jeopardizing the reliability of the entire system.

Many commercial computing platforms, including NVIDIA GPUs \cite{nvidia}, or ARM processors \cite{graviton,zena2}, integrate hardware-level ECCs based on Single-Error Correction and Double-Error Detection (SECDED) schemes to mitigate soft errors in memory subsystems \cite{sullivan2021characterizing}. While SECDED protects a subset of memory faults by correcting a single erroneous bit per memory line, its fault coverage remains inherently limited. Moreover, these mechanisms introduce non-negligible performance and energy overheads at runtime \cite{chen2018configurable}. Multiple research works have shown that the accuracy of CNNs and ViTs can be significantly influenced by faults in memory, using radiation \cite{badia2025reliability,bodmann2024evaluating,roquet2024cross} or simulation \cite{ahsaei2025lost,xue2023soft}. Recent experimental studies conducted under neutron irradiation \cite{roquet2024cross} demonstrate that enabling ECC on GPUs does not guarantee full protection against critical SDCs. These experiments reveal that ECCs reduce, but do not eliminate SDCs, and may even increase Detected Unrecoverable Error (DUE) rates, yet they do not prevent critical SDCs in CNN and ViT workloads.

A prior work comprehensively highlights the limitations of applying ECC to ViTs \cite{ahmadilivani2026LBR}. 
Multiple papers present alternative methods to conventional ECCs \cite{qin2017robustness, guan2019place, burel2021zero, lee2022value, ahmed2024nn, jo2025stegano, park2025pop, hong2024loco, huang2020functional, kim2024sparrow,traiola2023hardnning}. Among them, zero-space approaches apply error detection and correction mechanisms selectively to the important bits of CNNs' weights, by removing the Least Significant Bits (LSBs) and using them as a storage for parity to remove the ECCs' memory overhead. However, they do not consider ViTs, and also their protection capabilities are mostly limited to two or three error corrections. Moreover, \cite{ahsaei2025lost} proposes a bit-pattern–based protection scheme for parameters with float-32 datatype in ViTs. Nonetheless, its applicability across different ViT architectures and numerical formats remains limited. 


To the best of our knowledge, the interaction between ECC mechanisms and the fault resilience for both CNNs and ViTs has not been studied and compared. In this study, for the first time, we comprehensively study ECC's impact on such models, present new insights, especially for ViTs, and propose two novel lightweight alternatives to ECC, mostly outperforming it, with an open-source tool: 
\begin{itemize}
    \item Most Significant Exponent Triplication (MSET): selectively hardens the most vulnerable bits in CNN and ViT parameters, by triplicating the most significant exponent bit in floating point data types and storing its copies in the LSB positions.
    \item Chunk-wise Embedded Parity (CEP): uniformly protects fine-grain bit-chunks of values, each protected by one parity bit, embedded within the bits of a value, consistently providing stronger protection compared to SECDED.
\end{itemize}

Experimental results demonstrate that both methods significantly enhance the reliability of large CNN and ViT inference, with zero-space memory overhead, and considerably lower area and delay compared to SECDEC ECC. Overall, this work presents the first comprehensive study of ECC effectiveness on modern DL models and delivers substantial reliability enhancement for memory-intensive ML workloads operating in error-prone environments.

In the rest of the paper, Section \ref{sec:background} overviews the related works, Section \ref{sec:method} presents the methodology of MSET and CEP, Section \ref{sec:results} present the experimental implementation setup and results, Section \ref{sec:discussion} discuss the implications of the proposed methods and Section \ref{sec:conclusion} concludes the paper. 


\section{Related Works: ECC Alternatives for DNNs} \label{sec:background}

A range of alternative strategies to conventional ECCs has been explored to reduce their associated overheads. These approaches can be broadly classified into two categories. First, \textit{zero-space approaches} \cite{qin2017robustness, guan2019place, burel2021zero, lee2022value, ahmed2024nn, jo2025stegano, park2025pop, hong2024loco} eliminate the need for additional storage dedicated to parity bits by embedding protection within existing data representations. Second, \textit{selective ECC approaches} \cite{huang2020functional, kim2024sparrow, traiola2023hardnning} that reduce parity storage requirements by restricting ECC protection to a selected subset of bits that are considered more critical to model reliability.


In \textit{zero-space approaches}, the primary objective is to eliminate the additional storage overhead of ECC parity codes by embedding protection information into non-critical bits of weight representations. These methods exploit the limited numerical significance of certain bits, typically the LSBs, to incorporate error detection or correction capabilities directly within the data, thereby avoiding dedicated parity storage.
\textit{Weight Nulling} \cite{qin2017robustness} assumes that the LSB of a weight contributes negligibly to its numerical value and replaces it with a single parity bit for error detection; upon detecting an error, the affected weight is reset to zero. Similarly, \textit{Opportunistic Parity} \cite{burel2021zero} introduces an even-parity mechanism in which the LSB is substituted with a parity bit across multiple data formats, including 32-bit floating point (FP32), 16-bit Brain Floating Point (BF16), and 8-bit integer (INT8), with detected errors mitigated by zero masking. 

\textit{Value-Aware Parity Insertion} \cite{lee2022value} embeds two parity bits into weight representations based on their numerical value, focusing on weights within the range $[-0.5, 0.5]$ in an 8-bit binary format. This scheme encodes every 64-bit block, comprising eight weights, and enables double-error correction at the block level, albeit at the cost of a measurable reduction in model accuracy.
Stegano ECC \cite{jo2025stegano} applies a SEC(7,4) Hamming code to the most significant bits (MSBs) of FP32, FP16, and INT8 data types by partitioning each representation into segments of critical and non-critical bits. The resulting parity bits are embedded within the less significant portions of the same data word, allowing the correction of up to two faults per 32-bit data block without additional memory overhead.


\textit{LOCo} \cite{hong2024loco} protects compressed DNN weights by combining fine-grained in-DRAM SEC-DED ECC with multi-stage compression-aware structural checks. The ECC corrects one fault in 64-bit memory blocks storing compressed INT8 weights and detects multi-bit errors. \textit{PoP-ECC} \cite{park2025pop} extends this concept by proposing a two-tier memory protection scheme for CNNs targeting Multi-Bit Upsets (MBUs). It generates virtual parity bits that are not explicitly stored in memory; within a 64-bit block, two bits are removed from each 8-bit weight to accommodate 16 parity bits in the same block, enabling Reed–Solomon (RS)–based correction of up to two errors per block.

Some approaches embed parity information directly within weight values and rely on specialized training procedures to preserve model accuracy. \textit{In-place Zero-Space} \cite{guan2019place} exploits the observation that the most significant bit following the sign bit in INT8 representations is frequently zero, repurposing it to store parity bits for Hamming SEC–DED codes. To accommodate the presence of embedded parity bits, this method introduces a \textit{Weight Distribution-Oriented Training} strategy in which the network is trained with encoded weights. Similarly, \textit{NN-ECC} \cite{ahmed2024nn} integrates ECC parity bits directly into CNN parameters and employs a multi-task training framework designed to maintain inference accuracy while jointly learning the primary task and enforcing the embedded ECC constraints.


In the second category of approaches, selective ECC schemes protect only a subset of critical bits, typically the MSBs of a given data type. Huang et al. \cite{huang2020functional} employ a reinforcement learning–based framework to balance the trade-off between ECC redundancy overhead and model performance by selectively protecting MSBs. Their study demonstrates, however, that safeguarding only the MSBs is insufficient to ensure robust fault tolerance in CNNs, as errors occurring in least significant bits (LSBs) can also propagate and degrade inference accuracy.
\textit{Sparrow-ECC} \cite{kim2024sparrow} introduces a selective ECC mechanism for high-bandwidth memory that leverages value similarity in weight exponents to prevent fault-induced amplification of large weight values. The method protects critical bits by duplicating SEC codes, generating 128 bits of redundancy for each 1024-bit weight block, and constraining exponent values within a predefined expected range. By leveraging the capacity of off-the-shelf parity memory, \textit{Sparrow-ECC} aims to minimize additional storage overhead while enhancing reliability against memory faults. 

\textit{harDNNing} \cite{traiola2023hardnning} proposes a framework to identify and selectively protect critical DNN parameters and bits to improve fault tolerance. The approach first performs statistical fault injection on a subset of parameter bits to evaluate their impact on CNNs accuracy and label faults as critical or acceptable. These results are then used to train a machine learning classifier (i.e., Random Forest) that predicts the criticality of all parameters and bits across the network without exhaustive fault injection. Based on the predicted criticality, the framework applies selective Hamming ECC only to critical parameters or critical bits, instead of protecting the entire model. This selective protection significantly reduces memory overhead while maintaining reliability compared to conventional full ECC protection.

Despite the breadth of prior work, several important limitations remain unaddressed. While zero-space approaches are particularly attractive for modern large-scale DNNs due to their zero memory overhead and improved memory utilization, many existing solutions are limited to correcting two or three errors per block, even though each block may contain multiple weights and multiple errors. 
On the other hand, approaches that rely on training-time modifications alter model parameters and introduce significant complexity. 

In addition, the majority of studies focus exclusively on protecting weight values, neglecting other critical parameters within DNNs, and several approaches continue to depend on conventional ECC mechanisms that impose non-trivial complexity and redundancy overhead. Notably, existing methods have been evaluated extensively on CNNs, leaving their applicability to ViTs underexplored. To address these gaps, this work presents novel reliability mechanisms applicable to both CNNs and ViTs. We systematically analyze the impact of Hamming-based ECC, demonstrating its inherent limitations, and introduce new zero-space protection techniques that achieve high reliability while safeguarding multiple parameters without incurring ECC overhead.









\section{Methodology: Zero-Space Approaches for Parameter Protection} \label{sec:method}

\begin{figure}[h!]
    \includegraphics[width=0.3\textwidth]{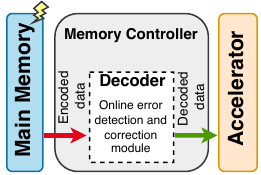}
    \centering
    \caption {An overall system view of the system for implementing the proposed zero-space mechanisms for error detection and correction. }
    
    \label{fig:system}
\end{figure}

This section presents two novel zero-space approaches proposed to protect all parameters of CNNs and ViTs against faults in the memories of DNN accelerators: \textit{MSET} and \textit{CEP}, respectively. Fig.~\ref{fig:system} illustrates an overall view of the system and how the proposed mechanisms perform error detection and correction in the execution flow. Data encoding is applied to the DNN's parameters in an offline mode, and the memory contains the encoded parameters prior to deployment. The encoded data in the memory, based on MSET or CEP, is decoded by the memory controller upon a read operation to detect and correct faults possibly occurring in the memory.

\subsection{Most Significant Exponent Triplication (MSET)}

\begin{figure}[h!]
    \vspace{-5mm}
    \includegraphics[width=0.45\textwidth]{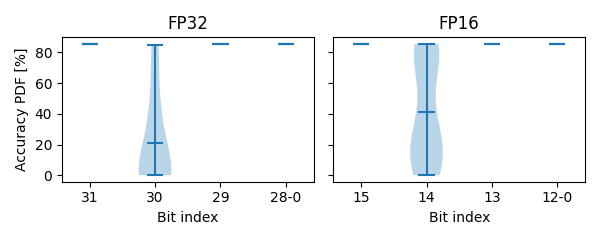}
    \centering
    \vspace{-2mm}
    \caption {Probability Distribution Function (PDF) of accuracy for ViT-base under 1000 repetitions of bit-wise fault injection into parameters.}
    \vspace{-5mm}
    \label{fig:bitwise}
\end{figure}

A bit-level resilience analysis, illustrated in Fig.~\ref{fig:bitwise}, reveals that the MSB of the exponent in both FP32 and FP16 representations is the most vulnerable to corruption, with even a single fault causing a substantial degradation in the inference accuracy of the DNN models. Based on this observation, we propose MSET, a bit-specific mechanism to protect critical bits in CNNs and ViTs. MSET leverages the fact that the two LSBs of the mantissa have a negligible impact on the numerical value of parameters and exhibit no measurable effect on model accuracy when perturbed.

\begin{figure}[h!]
    \includegraphics[width=0.45\textwidth]{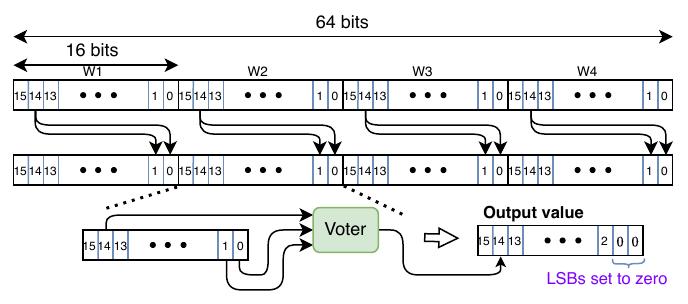}
    \centering
    \vspace{-2mm}
    \caption{Encoding and decoding the parameters in a memory block by the MSET method.}
    \vspace{-5mm}
    \label{fig:mset}
\end{figure}

In MSET, these two mantissa LSBs are repurposed to store two replicates of the exponent MSB for each parameter within a memory block, as shown in Fig.~\ref{fig:mset}. During decoding, the duplicated bits embedded in the mantissa LSBs are compared with the original exponent MSB using a majority-voting mechanism, and the resulting consensus bit is restored into the exponent MSB position. The two LSBs of the mantissa are then reset to $0$ in the reconstructed output value. This procedure is uniformly applied to all parameters within a memory block, which may comprise multiple parameters, providing fine-grained, zero-space protection against faults in the critical exponent bit.

\subsection{Chunk-wise Embedded Parity (CEP)}

\begin{figure}[h!]
    \includegraphics[width=0.45\textwidth]{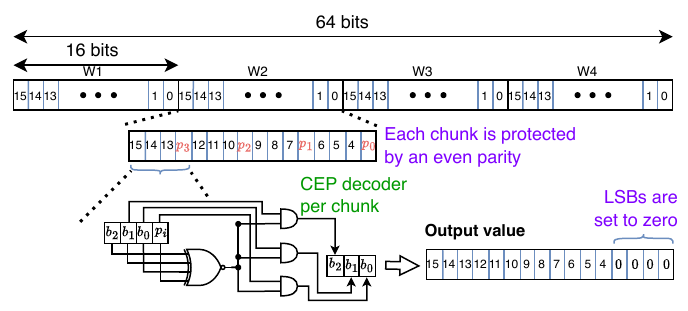}
    \centering
    \caption{Encoding and decoding the parameters in a memory block by the CEP method.}
    \label{fig:cep}
\end{figure}

As observed in Fig.~\ref{fig:bitwise}, the contribution of erroneous LSBs in both FP32 and FP16 representations to the accuracy degradation of CNNs and ViTs is minimal. This observation motivates the exploitation of multiple LSBs as an intrinsic redundancy source to protect more significant and accuracy-critical bits within model parameters. In this regard, we propose CEP to protect all bits in parameters in a fine-grained way through chunks of bits, as illustrated in Fig.~\ref{fig:cep}. The proposed CEP method partitions each parameter word into uniformly sized bit chunks, assigning one even-parity bit to each chunk. 

The parity bits are stored within the value itself by removing an equivalent number of LSBs, ensuring that all parity and data bits fit within the original bit-width of the data type. During inference, each embedded parity bit is interleaved with and checked against the chunk it protects. Upon detecting a parity mismatch, all bits in the corresponding chunk are reset to $0$, effectively mitigating the impact of corruption. The output value is then reconstructed by reordering the protected bits to their original positions while setting all LSBs to $0$. This fine-grained, chunk-level protection mechanism provides robust error detection and correction across all model parameters without requiring extra storage for parity bits.

\begin{figure}[h!]
    \centering
    \includegraphics[width=0.9\linewidth]{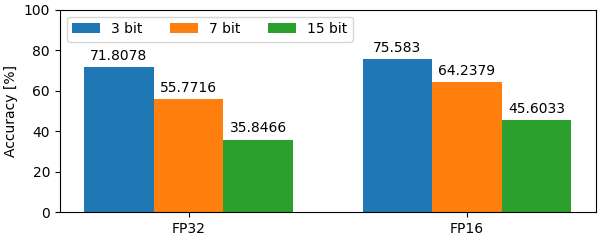}
    \caption{Chunk size exploration in CEP method for ViT-base under fault injection with BER=$3\times10^{-5}$.}
    \label{fig:chunk-exploration}
\end{figure}

To determine the optimal chunk size for the CEP scheme, we systematically evaluate all feasible chunk sizes for both FP32 and FP16 data types. As illustrated in Fig.~\ref{fig:chunk-exploration}, we consider only chunk configurations that uniformly partition the data word while accommodating the required parity bits within a data type. We conduct fault injection experiments on the parameters of the ViT-Base model, as a representative model, under a Bit Error Rate (BER) of $3\times10^{-5}$. The results indicate that the smallest evaluated chunk size, $3$ bits, consistently yields the highest level of protection across DNN parameters. Consequently, we select a chunk size of $3$ bits for CEP in both FP32 and FP16 representations.

\section{Experimental Results} \label{sec:results}

\subsection{Experimental Setup}

\subsubsection{Memory Model}

In this work, we focus on the main memory subsystem of an AI accelerator responsible for storing the parameters of a large-scale DNN. We consider 64-bit and 128-bit memory interfaces in the experiments to cover a broad range of configurations in edge-AI accelerators \cite{jouppi2017datacenter}. We evaluate two widely adopted numerical formats for parameters: FP32 and FP16, and analyze how their accommodation within memory lines interacts with error-correction mechanisms.
For ECC-based protection, each memory line is assumed to be protected using a standard SECDED Hamming ECC, with the corresponding check bits stored in dedicated parity memory. For elaboration, under a 64-bit memory line storing FP16 parameters, each line contains four parameters, meaning that ECC can correct at most one faulty parameter per four parameters. Using this memory model, we extract all parameters across a DNN model, including attention blocks, normalization, convolution, and FC layers, and apply our proposed protection techniques directly to these parameter sets.

\subsubsection{Fault Model and Fault Simulation}

For the memory fault model, we simulate soft errors by injecting random bit flips uniformly across both the parameter bits and ECC check bits. To capture the cumulative impact of memory faults, we conduct Fault Injection (FI) experiments across a range of Bit Error Rates (BERs). For each BER, we corrupt the model by introducing random bit flips at randomly selected bit positions throughout the entire parameter space prior to an inference. This procedure is repeated across multiple iterations to obtain statistically meaningful results. Iterations continue until the variation in the mean accuracy converges to within $1\%$, resulting in a repetition between $500$ and $1,500$ iterations, depending on the models, BER, and protection scheme. The resulting reliability metric is the average inference accuracy on the $4096$ images in the validation set over all iterations for each BER.

\subsubsection{Implementation}

We present the results on multiple representative large-scale DNNs with various architectures, including three CNNs (ResNet-152, MobileNet-V2, Inception) and three ViT models (ViT-base, DeiT-base, and Swin, with a patch size of 16), all pre-trained on the ImageNet-1K dataset. The FI experiments, encoding-decoding for ECC, MSET, and CEP, are implemented in Rust and integrated into a unified framework with PyTorch to load the CNNs and ViT models from Hugging Face. The methods are presented as a fully documented and reusable open-source framework named \textit{FaultForge}\footnote{\url{https://github.com/rezzubs/faultforge}} that can be adopted by researchers and engineers for similar studies. 
The core of the tool is a Python library for running FI experiments on DNN models with encoded parameters. 
It's designed in a modular way to enable the usage of custom encoders, test benches, and models.
It includes predefined configurations for the techniques presented in this paper, which are designed to be easily integrated with newly developed techniques. 
A key goal of the tool is to enable engineers and researchers to apply the techniques to their target DNNs and compare the results with alternative techniques using a shared API. 
A simple command-line interface is developed for performing experiments using the configurations covered in this paper. The experiments in this paper are performed on an NVIDIA A100 GPU with an AMD EPYC 7742 CPU. 


\subsection{Experimental Results}

\subsubsection{Impact of the Methods on the Accuracy}

Table~\ref{tab:accuracy} indicates the baseline accuracy of the evaluated models alongside their accuracy after applying the proposed MSET and CEP techniques to all parameters across different data types. Overall, both methods incur negligible accuracy loss. For ViTs in both FP32 and FP16, as well as CNNs in FP32, the accuracy degradation introduced by MSET and CEP is negligible, remaining identical in most cases and below $0.05\%$ in the worst-case. In CNNs operating in FP16, the impact remains limited ($<0.22\%$) for most architectures; however, MobileNet-V2 in FP16 exhibits a more noticeable degradation of approximately $1.5\%$. These results indicate that the proposed techniques preserve model accuracy in the vast majority of configurations.

The observed accuracy variations can be attributed to architectural differences among models. ViTs demonstrate higher resilience to small numerical perturbations due to their greater depth and larger parameter count, as well as the presence of attention mechanisms, normalization layers, and fully connected layers that can effectively mask minor errors. In contrast, CNNs rely heavily on convolutional layers, which offer fewer opportunities for error masking. MobileNet-V2 is particularly sensitive because it has the smallest parameter count among the evaluated models and employs depthwise separable convolutions, where a limited number of parameters are reused across many input activations. As a result, small perturbations can propagate more readily through the model. Since CEP repurposes multiple LSBs in the FP16 format, its effect on numerical precision is more pronounced in such architectures, explaining the relatively larger accuracy degradation observed in this particular case.

\begin{table}[h!]
\caption{Effect of MSET and CEP on the baseline accuracy of models.}
\resizebox{0.5\textwidth}{!}{
\begin{tabular}{|c|cc|cc|cc|}
\hline
Method       & \multicolumn{2}{c|}{\textbf{Baseline}}                   & \multicolumn{2}{c|}{\textbf{MSET}} & \multicolumn{2}{c|}{\textbf{CEP}} \\ \hline
Data Type       & \multicolumn{1}{c|}{FP32}    & \multicolumn{1}{c|}{FP16} & \multicolumn{1}{c|}{FP32}  & FP16  & \multicolumn{1}{c|}{FP32}  & FP16 \\ \hline
ViT-base        & \multicolumn{1}{c|}{85.50\%}        & \multicolumn{1}{c|}{85.50\%}     & \multicolumn{1}{c|}{85.50\%}      & 85.57\%      & \multicolumn{1}{c|}{85.50\%}      & 85.57\%     \\ \hline
DeiT-base       & \multicolumn{1}{c|}{81.79\%}        & \multicolumn{1}{c|}{81.76\%}     & \multicolumn{1}{c|}{81.79\%}      & 81.76\%      & \multicolumn{1}{c|}{81.79\%}      & 81.74\%     \\ \hline
Swin-Tiny       & \multicolumn{1}{c|}{81.40\%}        & \multicolumn{1}{c|}{81.40\%}     & \multicolumn{1}{c|}{81.40\%}      & 81.45\%      & \multicolumn{1}{c|}{81.40\%}      & 81.45\%     \\ \hline
ResNet-152      & \multicolumn{1}{c|}{82.28\%}        & \multicolumn{1}{c|}{82.30\%}     & \multicolumn{1}{c|}{82.28\%}      & 82.25\%      & \multicolumn{1}{c|}{82.28\%}      & 82.08\%     \\ \hline
MobileNet-V2    & \multicolumn{1}{c|}{72.39\%}        & \multicolumn{1}{c|}{72.46\%}     & \multicolumn{1}{c|}{72.39\%}      & 72.29\%      & \multicolumn{1}{c|}{72.39\%}      & 70.90\%     \\  \hline
Inception-V3    & \multicolumn{1}{c|}{76.76\%}        & \multicolumn{1}{c|}{72.68\%}     & \multicolumn{1}{c|}{76.71\%}      & 76.61\%      & \multicolumn{1}{c|}{76.73\%}      & 76.76\%     \\ \hline
\end{tabular}
}
\label{tab:accuracy}
\end{table}


\subsubsection{Reliability and Memory Overhead Analysis}

\begin{figure}[t!]
    \centering
    \includegraphics[width=0.5\textwidth]{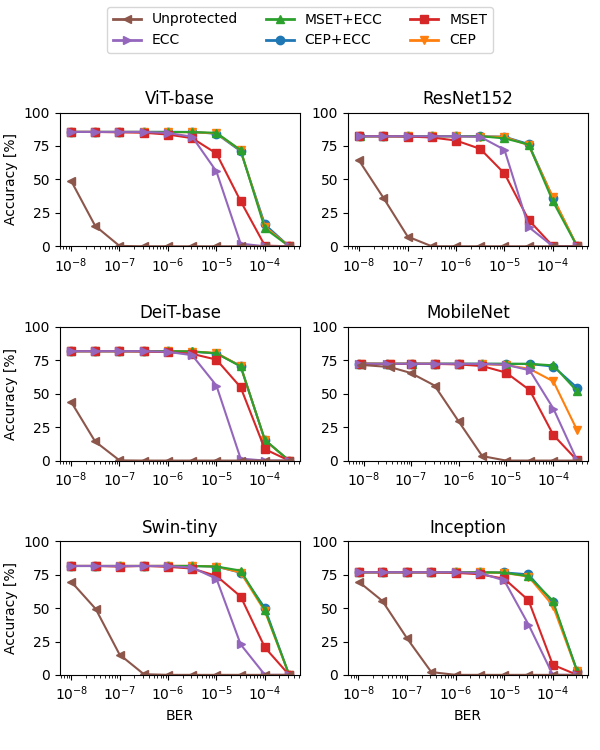}
    \vspace{-7mm}
    \caption{Reliability analysis of ViTs and CNNs using different protection mechanisms for FP32. }
    \label{fig:FI-fp32}
\end{figure}

\begin{figure}[t!]
    \centering
    \includegraphics[width=0.5\textwidth]{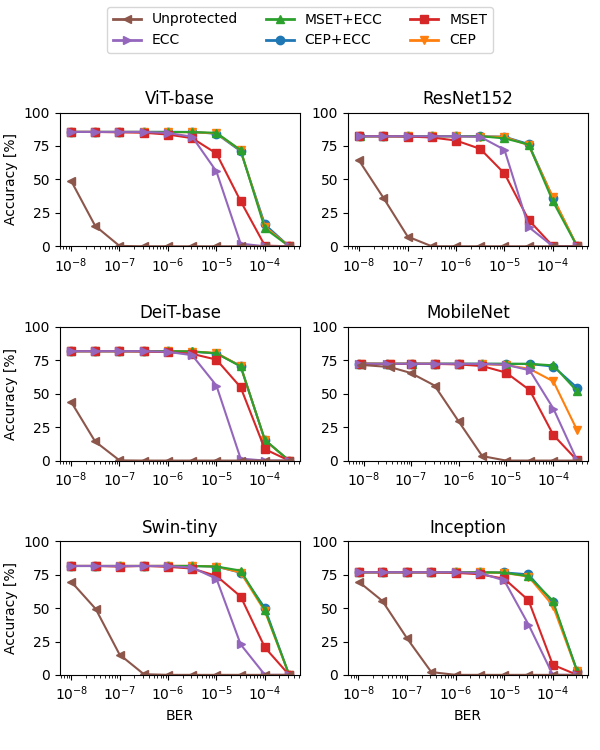}
    \vspace{-7mm}
    \caption{Reliability analysis of ViTs and CNNs using different protection mechanisms for FP16.}
    \label{fig:FI-fp16}
\end{figure}

Fig.~\ref{fig:FI-fp32} and Fig.~\ref{fig:FI-fp16} illustrate the reliability analysis of the proposed techniques in comparison with conventional ECC using a 64-bit memory block size for FP32 and FP16 representations, respectively. Across all evaluated models, the accuracy of unprotected DNNs degrades sharply even at relatively low BERs.
Notably, ViT models exhibit a more noticeable accuracy degradation than CNNs at the lowest BER (i.e., $10^{-8}$). This behavior can be attributed to the substantially larger number of parameters in ViTs, which increases the absolute number of erroneous bits under the same BER. Nevertheless, these results underscore that, for an identical BER, representing the fraction of bits corrupted in a given scenario, ViTs are inherently more susceptible to faults than CNNs, and their resilience can therefore be more severely compromised in the absence of effective protection.

Incorporating SECDED ECC substantially enhances resilience, preserving high model accuracy across BERS that are two to three orders of magnitude higher than those tolerated by unprotected CNN and ViT models (up to $10^{-5}$). Furthermore, the results indicate that smaller ECC-protected memory block sizes yield higher levels of protection, as fewer parameters are mapped to each memory line. Consequently, ECC covers a smaller group of parameters, increasing the error correction coverage. For clarity and readability, these additional results are not included in the figures.

However, the improved fault tolerance provided by ECC comes at the expense of substantial memory overhead. Specifically, the storage overhead introduced by ECC check bits is $12.5\%$ for 64-bit memory blocks and $7\%$ for 128-bit memory blocks. For instance, in the case of ViT-Base or DeiT-Base, which contain an identical number of parameters, deploying ECC in FP32 with 64-bit memory blocks requires an additional $345.6$~MB of memory to store check bits, or, FP16 with 128-bit memory blocks incurs an extra $96.8$~MB. Such significant memory overheads are non-trivial for large-scale DNNs and pose considerable challenges for both edge and cloud platforms, where memory capacity, bandwidth, and energy efficiency are critical constraints.

The results demonstrate that the proposed MSET and CEP techniques provide substantially stronger protection for both ViTs and CNNs than conventional ECC. 
Across all evaluated BERs, the average accuracy of ViTs
protected with both MSET or CEP consistently exceeds that of ECC-protected models, with the advantage becoming more pronounced at higher BERs. For CNNs, however, MSET demonstrates marginally lower protection than ECC in some cases. These results highlight that ViTs specifically can achieve high resilience through merely protecting their most significant exponent bit. 
This finding suggests that lightweight, zero-overhead mechanisms such as MSET can enable highly reliable deployment without incurring the complexity and storage overhead of ECC. Nevertheless, since other bits remain unprotected, faults affecting less significant bits begin to degrade accuracy at higher BERs. In such cases, MSET can be combined with ECC to further strengthen protection.

CEP, in contrast, provides an even stronger protection mechanism for all DNNs. In most experiments, CEP achieves resilience comparable to that of the combined MSET+ECC approach, while the addition of ECC on top of CEP yields no further measurable improvement. These results indicate that CEP can serve as a strong alternative to traditional ECC, effectively eliminating the need for ECC in CNNs while achieving very high reliability for both CNNs and ViTs. Notably, across all CEP evaluations, models remain functional at BERs ranging from $3\times10^{-5}$ to $10^{-4}$, corresponding to tens of thousands of faults within model parameters, underscoring the effectiveness of CEP under severe fault conditions. Therefore, CEP can improve the resilience of DNNs by one order of magnitude higher than that of ECC.

\subsubsection{Hardware Overhead Analysis}

To assess the hardware overhead of the proposed techniques at the memory controller, we implement the decoders for MSET and CEP, Hamming SECDED in VHDL. All designs are synthesized using Cadence Genus synthesis tool using Cadence’s free 45nm PDK’s technology library \textit{fast\_vdd1v0\_basicCells.lib}. Table~\ref{tab:hardware} reports the area and delay comparisons across different protection mechanism configurations. Since MSET is data-type dependent, we implement separate decoders for FP16 and FP32. In contrast, both ECC and CEP are agnostic to data type, and their implementations are determined solely by the memory block width.

As shown in Table~\ref{tab:hardware}, conventional ECC incurs the highest area and delay overhead, with both metrics increasing as the memory block size grows due to the complexity of the decoding logic. MSET exhibits the lowest area and delay among all evaluated schemes, reflecting its simple detection and correction mechanism. Its delay remains constant at $35 ps$ across all configurations, making it approximately $15\times$ and $20.5\times$ faster than ECC for 64-bit and 128-bit blocks, respectively. While the area of MSET scales with the data type or block size due to its dependence on them, its delay remains unaffected, as it decodes each value in parallel. 

On the other hand, CEP is independent of data type and provides significantly lower overheads than ECC while delivering the strongest protection. Its delay is independent of block size, as all chunks are decoded in parallel, and its area scales proportionally with block size. Compared to ECC, CEP achieves an area reduction of approximately $3.5\times$ for 64-bit blocks and $3.3\times$ for 128-bit blocks. Moreover, CEP's critical path is nearly $5\times$ shorter than that of 64-bit SECDED and nearly $7\times$ shorter than that of 128-bit SECDED, respectively. This indicates that decoders based on both MSET and CEP are significantly more lightweight in terms of hardware overhead than SECDED.

\begin{table}[t!]
\caption{Hardware synthesis results for the decoder of different methods.}
\label{tab:hardware}
\resizebox{0.5\textwidth}{!}{
\begin{tabular}{|c|cc|cc|}
\hline
\begin{tabular}[c]{@{}c@{}}memory line \\ bit-width\end{tabular} & \multicolumn{2}{c|}{64-bit}                  & \multicolumn{2}{c|}{128-bit}                 \\ \hline
                                                                 & \multicolumn{1}{c|}{Delay ($ps$)} & Area ($\mu m^2$) & \multicolumn{1}{c|}{Delay ($ps$)} & Area ($\mu m^2$) \\ \hline
SECDED Hamming                                                  & \multicolumn{1}{c|}{526}        & 631.67     & \multicolumn{1}{c|}{720}        & 1189.13    \\ \hline
MSET FP16                                                        & \multicolumn{1}{c|}{35}         & 13.68      & \multicolumn{1}{c|}{35}         & 27.36      \\ \hline
MSET FP32                                                        & \multicolumn{1}{c|}{35}         & 6.84       & \multicolumn{1}{c|}{35}         & 13.68      \\ \hline
CEP (3-bit chunks)                                               & \multicolumn{1}{c|}{108}        & 180.58     & \multicolumn{1}{c|}{108}        & 361.16     \\ \hline
\end{tabular}
}
\end{table}

\subsubsection{Comparison with SOTA}

\begin{table*}[t!]
\caption{Comparison between CEP (our strongest proposed method) with zero-space state-of-the-art works that include hardware implementation.}
\label{tab:comparison}
\centering
\begin{tabular}{|c|c|c|c|c|}
\hline
Method & Stegano-ECC \cite{jo2025stegano} & PoP-ECC \cite{park2025pop}  & LOCo \cite{hong2024loco}    & \textbf{Proposed CEP}                                                               \\ \hline

Evaluated Models      & CNNs and ViT-base  & CNNs  & \begin{tabular}[c]{@{}c@{}}CNNs and \\ BERT\end{tabular}       & \begin{tabular}[c]{@{}c@{}}\textbf{CNNs and }\\ \textbf{multiple ViTs}\end{tabular} \\ \hline

\begin{tabular}[c]{@{}c@{}}Protection \\ Capability\end{tabular} & \begin{tabular}[c]{@{}c@{}}3-bit detection and \\2-bit correction \\ in 32 bits\end{tabular} &\begin{tabular}[c]{@{}c@{}}3-bit detection and \\ 2-bit correction and \\ in 64 bits\end{tabular} & \begin{tabular}[c]{@{}c@{}}2-bit detection and \\ 1-bit correction \\ in 64 bits\end{tabular}  & \begin{tabular}[c]{@{}c@{}}\textbf{Simultaneous error detection} \\ \textbf{and mitigation for 16 chunks} \\ \textbf{in 64 bits}\end{tabular}   \\ \hline

\begin{tabular}[c]{@{}c@{}}Training \\ Requirement\end{tabular}  & No     & No  & No    & \textbf{No}   \\ \hline

Data Types  & \begin{tabular}[c]{@{}c@{}}FP32, FP16, \\ INT8\end{tabular}    & INT8      & \begin{tabular}[c]{@{}c@{}}FP32, FP16, \\ INT8\end{tabular}    & \textbf{FP32, FP16}  \\ \hline
\begin{tabular}[c]{@{}c@{}}Technology \\ ($nm$)\end{tabular} & 7 & 28  & 32    & \textbf{45} \\ \hline
Area ($\mu m^2$) & 1000  & 1760  & 18900  & \textbf{180.58}  \\ \hline
\end{tabular}
\end{table*}

Table~\ref{tab:comparison} compares CEP (the strongest of the proposed techniques) with recent state-of-the-art zero-space protection methods. Unlike prior approaches, CEP is evaluated across multiple large-scale CNN and ViT models and is shown to be a robust alternative to conventional ECC for a wide range of DNN architectures. CEP provides broader protection coverage by partitioning each memory block into fine-grained 3-bit chunks; for example, in a 64-bit block, it can detect and mitigate errors in up to 16 individual chunks simultaneously. The method operates without requiring retraining of DNN models or additional memory for parity storage and is agnostic to data type, supporting both FP32 and FP16 representations. Moreover, as summarized in the table, the CEP decoder incurs the lowest hardware area overhead among all compared methods, further highlighting its efficiency and practicality.

\section{Discussion}  \label{sec:discussion}

This work introduces MSET and CEP, two lightweight, zero-space error detection and correction techniques, and demonstrates their effectiveness across a wide range of CNN and ViT architectures. The experiments show that both methods provide strong alternatives to conventional ECC by delivering high resilience against memory faults with minimal impact on model accuracy. Unlike ECC, MSET and CEP require no additional storage for parity bits and only modest hardware modifications in the memory decoder, enabling efficient, run-time fault correction. These properties make the proposed techniques particularly attractive for modern large-scale DNN workloads, where memory overhead and performance constraints are critical.

An important observation from our evaluation is that ViTs benefit more significantly from MSET and CEP than CNNs. Due to their substantially larger number of parameters and deeper architectures, ViTs are inherently more vulnerable to memory faults and therefore gain greater reliability improvements from zero-space protection mechanisms. Moreover, the proposed techniques remain effective across both low and high BERs, highlighting their applicability to diverse environments ranging from supercomputers to edge systems. In particular, CEP consistently achieves high resilience without relying on ECC, positioning it as a practical replacement for ECC in industrial deployments.

In the paper, MSET and CEP are presented as post-training protection mechanisms. For safety-critical applications, such as autonomous vehicles, rigorous safety analysis is required to detect and mitigate rare corner cases that may arise during deployment. Nonetheless, these techniques can also be integrated into the training process, allowing the model to adapt to the modified bit representations and potentially eliminating any accuracy degradation while addressing safety concerns at run time. Such integration represents a promising direction for deploying these methods in highly critical environments.

Finally, even in scenarios where system designers prefer not to embed parity information directly within parameter bits, the proposed techniques still offer advantages over traditional ECC. The required additional memory remains comparable to or lower than ECC overheads, while incurring smaller area and performance costs in the decoder and achieving higher reliability. Furthermore, in the case of CEP, memory overhead can be further reduced by increasing the chunk size, enabling flexible trade-offs between protection strength and resource utilization. Overall, these properties make MSET and CEP compelling solutions for reliable and scalable deployment of modern DNN workloads at both the edge and in the HPCs.
\section{Conclusions} \label{sec:conclusion}

This paper presents two alternative zero-space approaches to ECCs for large-scale DNNs. Comprehensively validated on multiple CNNs and ViT architectures, it is shown that the proposed MSET and CEP methods can achieve similar or higher resilience in DNNs with significantly lower overhead. Results indicate that ViTs can be effectively protected by merely protecting their highest exponent bits in FP16 and FP32 representations. Furthermore, applying the CEP technique can guarantee the resilience of DNNs by up to one order of magnitude higher BERs, with a $3.5\times$ lower area overhead and $7\times$ faster decoder compared to SECDED ECC.  Overall, this work presents the first comprehensive study of ECC effectiveness on modern DL models and presents novel alternative techniques, achieving substantial reliability gains for memory-intensive DL workloads operating in error-prone environments.

\section*{Acknowledged} 

This paper is supported in part by  EU Grant Project 101160182 “TAICHIP”, and the EU Grant 101194287 “NexTArc”.

\bibliographystyle{IEEEtran}
\bibliography{refs.bib}
\end{document}